\documentclass[aps,preprint]{revtex4}%
\usepackage{amsfonts}
\usepackage{amsmath}
\usepackage{amssymb}
\usepackage{graphicx}%
\providecommand{\U}[1]{\protect\rule{.1in}{.1in}}

\begin{document}
\preprint{ }
\title[excitation exchange]{Excitation exchange between identical atoms having arbitrary angular momentum}
\thanks{Contact: P. R. Berman, pberman@umich.edu}
\author{P. R. Berman}
\affiliation{Physics Department, University of Michigan, Ann Arbor, MI \ 48109-1040}
\author{Zeyuan Wang}
\affiliation{Physics Department, University of Michigan, Ann Arbor, MI \ 48109-1040}
\begin{abstract}
Equations of motion are derived for excitation exchange between two atoms
having arbitrary angular momentum.

\end{abstract}
\volumeyear{year}
\volumenumber{number}
\issuenumber{number}
\eid{identifier}
\date[Date text]{date}
\startpage{1}
\maketitle

\section{\bigskip Excitation Exchange}

The exchange of excitation between two identical atoms represents a
fundamental problem in quantum optics. The original paper that treated this
exchange process was written by Stephen \cite{st}, but did not contain
explicit evaluations for states of definite angular momentum. Hutchinson and
Hameka \cite{ham} extended Stephen's calculation to atoms having a ground state
with total angular momentum $J=0$ and an excited state with angular momentum
$J=1$ for atoms on the $z-$axis, neglecting any effects of retardation in the
excitation exchange (Milonni and Knight \cite{mk} repeated the calculation,
including effects of retardation). General expressions of excitation exchange
neglecting retardation for $J=0-1$ transitions can be found in the article by
Fu and Berman \cite{fu} for arbitrary orientations of the atoms. If the
calculations are extended to states having arbitrary angular momentum, new
physical features can appear. In particular, for atoms on the $z-$axis, there
can be a "swap" of dipole optical coherence that is not possible for $J=0-1$
transitions. In this note we extend the excitation exchange results to states
having arbitrary angular momentum, neglecting retardation. For most atomic
systems of practical importance, the effects of retardation in the excitation
exchange process are negligible.

We consider two identical atoms having ground states with angular momentum $G$
and excited states with angular momentum $H$. The ground state kets are
denoted by $\left\vert Gm_{G}\right\rangle $ and the excited state kets by
$\left\vert Hm_{H}\right\rangle $, where $m_{G}$ and $m_{H}$ label the
magnetic state sublevels with $m_{G}$ running from $-G$ to $G$ and $m_{H}$
running from $-H$ to $H$. Atom 1 is located at the origin and atom 2 at
position $\mathbf{R}_{2.}$ The calculation can be extended easily to any
number of atoms, but the number of coupled equations increases rapidly with
increaded atom number.

In dipole approximation and the rotating-wave-approximation (RWA), the
interaction Hamiltonian $V$ for the atom-field system can be written as%
\begin{equation}
V=-\sum_{j=1}^{2}\left[  \boldsymbol{\mu}_{+}^{(j)}\cdot\mathbf{E}_{+}\left(
\mathbf{R}_{j}\right)  +\mathbf{E}_{-}\left(  \mathbf{R}_{j}\right)
\cdot\boldsymbol{\mu}_{-}^{(j)}\right]  ,\label{hrwa}%
\end{equation}
where
\begin{align}
\boldsymbol{\mu}_{+}^{(j)} &  =\sum_{m_{G},m_{H}}\left[  \boldsymbol{\mu}%
_{-}\left(  Gm_{G},Hm_{H}\right)  \right]  ^{\ast}\left(  \left\vert
Hm_{H}\right\rangle \left\langle Gm_{G}\right\vert \right)  _{j},\\
\boldsymbol{\mu}_{-}^{(j)} &  =\sum_{m_{G},m_{H}}\boldsymbol{\mu}_{-}\left(
Gm_{G},Hm_{H}\right)  \left(  \left\vert Gm_{G}\right\rangle \left\langle
Hm_{H}\right\vert \right)  _{j},
\end{align}
$\left(  \left\vert Hm_{H}\right\rangle \left\langle Gm_{G}\right\vert
\right)  _{j}$ is a raising operator for atom $j$, $\left(  \left\vert
Gm_{G}\right\rangle \left\langle Hm_{H}\right\vert \right)  _{j}$ is a
lowering operator for atom $j$, $\boldsymbol{\mu}_{-}\left(  Gm_{G}%
,Hm_{H}\right)  $ is a dipole moment matrix element between states
$\left\langle Gm_{G}\right\vert $ and $\left\vert Hm_{H}\right\rangle $,
\begin{equation}
\mathbf{E}_{+}\left(  \mathbf{R}\right)  =\frac{i}{\left(  2\pi\right)
^{3/2}}\sum_{\lambda=1,2}\int d\mathbf{k}\left(  \frac{\hbar\omega_{k}%
}{2\epsilon_{0}}\right)  ^{1/2}a\left(  \mathbf{k}_{\lambda}\right)
\boldsymbol{\epsilon}_{\mathbf{k}}^{(\lambda)}e^{i\mathbf{k\cdot R}}=\left[
\mathbf{E}_{-}\left(  \mathbf{R}\right)  \right]  ^{\dag}%
\end{equation}
is the positive component of the electric field operator, $a\left(
\mathbf{k}_{\lambda}\right)  $ is an annihilation operator for a field mode
having wave vector%
\begin{equation}
\mathbf{k}=k\left(  \sin\theta_{k}\cos\phi_{k}\mathbf{\hat{x}}+\sin\theta
_{k}\sin\phi_{k}\mathbf{\hat{y}+}\cos\theta_{k}\mathbf{\hat{z}}\right)  ,
\end{equation}
frequency $\omega_{k}=kc,$ and polarization $\boldsymbol{\epsilon}%
_{\mathbf{k}}^{(\lambda)}$ $\left(  \lambda=1,2\right)  ,$ with
\begin{subequations}
\label{ep}%
\begin{align}
\boldsymbol{\epsilon}_{\mathbf{k}}^{(1)} &  =\boldsymbol{\epsilon}_{\mathbf{k}}^{(\theta_{k})}=\cos\theta_{k}\cos
\phi_{k}\mathbf{\hat{x}}+\cos\theta_{k}\sin\phi_{k}\mathbf{\hat{y}}-\sin
\theta_{k}\mathbf{\hat{z},}\\
\boldsymbol{\epsilon}_{\mathbf{k}}^{(2)} &  =\boldsymbol{\epsilon}_{\mathbf{k}}^{(\phi_{k})}=-\sin\phi_{k}%
\mathbf{\hat{x}}+\cos\phi_{k}\mathbf{\hat{y},}
\end{align}
Note that we can write
\end{subequations}
\begin{equation}
\boldsymbol{\mu}_{-}\left(  Gm_{G},Hm_{H}\right)  \cdot\boldsymbol{\epsilon
}_{\mathbf{k}}^{(\lambda)}=\mu_{GH}M_{m_{G},m_{H}}^{(\lambda)}\left(
\Omega_{k}\right)  ,
\end{equation}
where $\mu_{GH}$ is a reduced matrix element (assumed real), $\Omega_{k}$ is a
solid angle, and
\begin{equation}
M_{m_{G},m_{H}}^{(\lambda)}=\frac{1}{\sqrt{2G+1}}\sum_{q=-1}^{1}%
\begin{bmatrix}
H & 1 & G\\
m_{H} & q & m_{G}%
\end{bmatrix}
(-1)^{q}\boldsymbol{\epsilon}_{-q}^{(\lambda)}\left(  \Omega_{k}\right)
\end{equation}
where%
\begin{subequations}
\begin{align}
\boldsymbol{\epsilon}_{\pm1}^{(\lambda)}\left(  \Omega_{k}\right)   &
=\mp\frac{\left[  \boldsymbol{\epsilon}_{\mathbf{k}}^{(\lambda)}\right]
_{x}\pm i\left[  \boldsymbol{\epsilon}_{\mathbf{k}}^{(\lambda)}\right]  _{y}%
}{\sqrt{2}},\\
\boldsymbol{\epsilon}_{0}^{(\lambda)}\left(  \Omega_{k}\right)   &  =\left[
\boldsymbol{\epsilon}_{\mathbf{k}}^{(\lambda)}\right]  _{z},
\end{align}
and the quantity in square brackets is a Clebsch-Gordan coefficient.

In the absence of any input fields and with at most one excitation of the two
atoms, the state vector for the two-atom system can be written in an
interaction representation as%

\end{subequations}
\begin{align}
\left\vert \psi(t)\right\rangle  &  =\sum_{m_{G},m_{H}}b_{m_{H},m_{G}%
;0}(t)e^{-i\left(  \omega_{m_{G}}+\omega_{m_{H}}\right)  t}\left\vert
m_{H},m_{G};0\right\rangle \nonumber\\
&  +\sum_{m_{G},m_{H}}b_{m_{G},m_{H};0}(t)e^{-i\left(  \omega_{m_{G}}%
+\omega_{m_{H}}\right)  t}\left\vert m_{G},m_{H};0\right\rangle \nonumber\\
&  +\sum_{m_{G},m_{G}^\prime}\sum_{\lambda=1,2}\int d\mathbf{k}b_{m_{G},m_{G}^{\prime}}^{(\lambda
)}(\mathbf{k},t)e^{-i\left(  \omega_{m_{G}}+\omega_{m_{G}^{\prime}}\right)
t}e^{-i\omega_{k}t}\left\vert m_{G},m_{G}^{\prime};\mathbf{k}_{\lambda
}\right\rangle ,
\end{align}
where $b_{m_{H},m_{G};0}(t)$ is the state amplitude for atom 1 to be in its
$Hm_{H}$ state, atom 2 to be in its $Gm_{G}$ state, and the field to be in the
vacuum state at time $t,$ $b_{m_{G},m_{H};0}(t)$ is the state amplitude for
atom 1 to be in its $Gm_{G}$ state, atom 2 to be in its $Hm_{H}$ state, and
the field to be in the vacuum state at time $t$, and $b_{m_{G},m_{G}^{\prime}%
}^{(\lambda)}(\mathbf{k},t)$ is the state amplitude for atom 1 to be in its
$Gm_{G}$ state, atom 2 to be in its $Gm_{G}^{\prime}$ state, and the field to
be in state $\left\vert \mathbf{k}_{\lambda}\right\rangle $ at time $t$. The
frequencies $\omega_{m_{G}}$ and $\omega_{m_{H}}$ refer to ground and excited
state frequencies in the presence of a magnetic field along the $z-$axis.

In an interaction representation, the state amplitudes evolve as
\begin{subequations}
\begin{align}
\dot{b}_{m_{H},m_{G};0}(t) &  =-\frac{\mu_{GH}}{\left(  2\pi\right)  ^{3/2}%
}\sum_{m_{G}^{\prime}}\sum_{\lambda=1,2}\int d\mathbf{k}\left(  \frac
{\omega_{k}}{2\hbar\epsilon_{0}}\right)  ^{1/2}\left[  M_{m_{G}^{\prime}%
,m_{H}}^{(\lambda)}\left(  \Omega_{k}\right)  \right]  ^{\ast}\nonumber\\
&  \times e^{i\left(  \omega_{m_{H},m_{G}^{\prime}}-\omega_{k}\right)
t}b_{m_{G}^{\prime},m_{G}}^{(\lambda)}(\mathbf{k},t),\\
\dot{b}_{m_{G},m_{H};0}(t) &  =-\frac{\mu_{GH}}{\left(  2\pi\right)  ^{3/2}%
}\sum_{m_{G}^{\prime}}\sum_{\lambda=1,2}\int d\mathbf{k}\left(  \frac
{\omega_{k}}{2\hbar\epsilon_{0}}\right)  ^{1/2}\left[  M_{m_{G}^{\prime}%
,m_{H}}^{(\lambda)}\left(  \Omega_{k}\right)  \right]  ^{\ast}%
e^{i\mathbf{k\cdot R}_{2}}\nonumber\\
&  \times e^{i\left(  \omega_{m_{H},m_{G}^{\prime}}-\omega_{k}\right)
t}b_{m_{G},m_{G}^{\prime}}^{(\lambda)}(\mathbf{k},t),\\
\dot{b}_{m_{G}^{\prime},m_{G}}^{(\lambda)}(\mathbf{k},t) &  =\frac{\mu_{GH}%
}{\left(  2\pi\right)  ^{3/2}}\sum_{m_{H}^{\prime}}\left(  \frac{\omega_{k}%
}{2\hbar\epsilon_{0}}\right)  ^{1/2}M_{m_{G}^{\prime},m_{H}^{\prime}%
}^{(\lambda)}\left(  \Omega_{k}\right)  e^{-i\left(  \omega_{m_{H}^{\prime
},m_{G}^{\prime}}-\omega_{k}\right)  t}b_{m_{H}^{\prime},m_{G};0}%
(t)\nonumber\\
&  +\frac{\mu_{GH}}{\left(  2\pi\right)  ^{3/2}}\sum_{m_{H}^{\prime}}\left(
\frac{\omega_{k}}{2\hbar\epsilon_{0}}\right)  ^{1/2}M_{m_{G},m_{H}^{\prime}%
}^{(\lambda)}\left(  \Omega_{k}\right)  e^{-i\mathbf{k\cdot R}_{2}%
}e^{-i\left(  \omega_{m_{H}^{\prime},m_{G}}-\omega_{k}\right)  t}%
b_{m_{G}^{\prime},m_{H}^{\prime};0}(t),
\end{align}
where $\omega_{\alpha,\alpha^{\prime}}=\omega_{\alpha}-\omega_{\alpha^{\prime
}}$.

The equation for $\dot{b}_{m_{G}^{\prime},m_{G}}^{(\lambda)}(\mathbf{k},t)$ is
formally integrated over time and substituted into the equations for $\dot
{b}_{m_{H},m_{G};0}(t)$ and $\dot{b}_{m_{G},m_{H};0}(t)$ and the integrations
and summations are carried out. In the Weisskopf-Wigner approximation (WWA)
\cite{ww}, the $\mathbf{R}_{2}$-independent terms simply give rise to
spontaneous decay, whereas the $\mathbf{R}_{2}$-dependent terms are
responsible for excitation exchange. The spontaneous emission rate $\gamma
_{H}=2\gamma$ is given by%
\end{subequations}
\begin{equation}
\gamma_{H}=\frac{\mu_{GH}^{2}\omega_{0}^{3}}{3\left(  2H+1\right)  \pi
\epsilon_{0}\hbar c^{3}},
\end{equation}
where $\omega_{0}$ is the transition frequency in the absence of any external
magnetic field. If we neglect retardation in the excitation exchange (that is
we assume that $\gamma_{H}R_{2}/c\ll1$, $\Delta_{H}R_{2}/c\ll1$ and $\omega_{B}R_{2}/c\ll1$, where $\Delta_{H}$ is a level shift associated with excitation exchange and
$\omega_{B}$ is a Zeeman shift), we find
\begin{subequations}
\label{ev}%
\begin{align}
\dot{b}_{m_{H},m_{G};0}(t) &  =-\gamma b_{m_{H},m_{G};0}(t)-\gamma\sum
_{m_{G}^{\prime},m_{H}^{\prime}}G_{m_{H},m_{G}}^{m_{G}^{\prime},m_{H}^{\prime
}}\left(  \mathbf{R}_{2},t\right)  b_{m_{G}^{\prime},m_{H}^{\prime};0}(t),\\
\dot{b}_{m_{G},m_{H};0}(t) &  =-\gamma b_{m_{G},m_{H};0}(t)-\gamma\sum
_{m_{G}^{\prime},m_{H}^{\prime}}G_{m_{G},m_{H}}^{m_{H}^{\prime},m_{G}^{\prime
}}\left(  \mathbf{R}_{2},t\right)  b_{m_{H}^{\prime},m_{G}^{\prime};0}(t),
\end{align}
where
\end{subequations}
\begin{subequations}
\begin{gather}
G_{m_{H},m_{G}}^{m_{G}^{\prime},m_{H}^{\prime}}\left(  \mathbf{R}_{2},t\right)
=\frac{\mu_{GH}^{2}}{\gamma\left(  2\pi\right)  ^{3}}\left(  \frac{\omega
_{0}^{3}}{2\hbar\epsilon_{0}c^{3}}\right)  \sum_{\lambda=1,2}\int d\omega
_{k}d\Omega_{k}\int_{0}^{t}dt^{\prime}e^{i\left(  \omega_{m_{H},m_{G}^{\prime
}}t-\omega_{m_{H}^{\prime},m_{G}}t^{\prime}\right)  }\nonumber\\
\times e^{-i\omega_{k}\left(  t-t^{\prime}\right)  }\left[  M_{m_{G}^{\prime
},m_{H}}^{(\lambda)}\left(  \Omega_{k}\right)  \right]  ^{\ast}M_{m_{G}%
,m_{H}^{\prime}}^{(\lambda)}\left(  \Omega_{k}\right)  e^{-i\mathbf{k\cdot
R}_{2}},\\
G_{m_{G},m_{H}}^{m_{H}^{\prime},m_{G}^{\prime}}\left(  \mathbf{R}_{2},t\right)
=\frac{\mu_{GH}^{2}}{\gamma\left(  2\pi\right)  ^{3}}\left(  \frac{\omega
_{0}^{3}}{2\hbar\epsilon_{0}c^{3}}\right)  \sum_{\lambda=1,2}\int d\omega
_{k}d\Omega_{k}\int_{0}^{t}dt^{\prime}e^{i\left(  \omega_{m_{H},m_{G}^{\prime
}}t-\omega_{m_{H}^{\prime},m_{G}}t^{\prime}\right)  }\nonumber\\
\times e^{-i\omega_{k}\left(  t-t^{\prime}\right)  }\left[  M_{m_{G}^{\prime
},m_{H}}^{(\lambda)}\left(  \Omega_{k}\right)  \right]  ^{\ast}M_{m_{G}%
,m_{H}^{\prime}}^{(\lambda)}\left(  \Omega_{k}\right)  e^{i\mathbf{k\cdot
R}_{2}}=G_{m_{H},m_{G}}^{m_{G}^{\prime},m_{H}^{\prime}}\left(  -\mathbf{R}%
_{2},t\right)  .
\end{gather}

We expand
\end{subequations}
\[
e^{-i\mathbf{k\cdot R}_{2}}=4\pi\sum_{\ell,m}(-i)^{\ell}Y_{\ell m}\left(
\Omega_{k}\right)  \left[  Y_{\ell m}\left(  \mathbf{\hat{R}}_{2}\right)
\right]  ^{\ast}j_{\ell}(kR_{2}),
\]
where $j_{\ell}$ is a spherical Bessel function and $Y_{\ell m}$ is a spherical
harmonic. In WWA, we replace $j_{\ell}(kR_{2})$ by $j_{\ell}\left[  k_{0}%
R_{2}+\left(  k-k_{0}\right)  R_{2}\right]  $ and keep terms involving
$\left(  k-k_{0}\right)  R_{2}$ only when they appear in phases. We then write%
\[
j_{\ell}=\frac{h_{\ell}+h_{\ell}^{\left(  2\right)  }}{2},
\]
where $h_{\ell}$ and $h_{\ell}^{\left(  2\right)  }$ are spherical Hankel
functions of the first and second kind, respectively, and integrate over
$\omega_{k}$ and $t^\prime$ to arrive at
\begin{align}
G_{m_{H},m_{G}}^{m_{G}^{\prime},m_{H}^{\prime}}\left(  \mathbf{R}_{2},tw\right)
&  =\frac{4\pi^{2}\mu_{GH}^{2}}{\gamma\left(  2\pi\right)  ^{3}}\left(
\frac{\omega_{0}^{3}}{2\hbar\epsilon_{0}c^{3}}\right)  \sum_{\lambda=1,2}\int
d\Omega_{k}e^{i\left(  \omega_{m_{H},m_{G}^{\prime}}-\omega_{m_{H}^{\prime
},m_{G}}\right)  t}\left[  M_{m_{G}^{\prime},m_{H}}^{(\lambda)}\left(
\Omega_{k}\right)  \right]  ^{\ast}\nonumber\\
&  \times M_{m_{G},m_{H}^{\prime}}^{(\lambda)}\left(  \Omega_{k}\right)
\sum_{\ell,m}(-i)^{\ell}Y_{\ell m}\left(  \Omega_{k}\right)  \left[  Y_{\ell
m}\left(  \mathbf{\hat{R}}_{2}\right)  \right]  ^{\ast}h_{\ell}(k_{0}%
R_{2})\nonumber\\
&  =\frac{3}{2}\left(  2H+1\right)  \sum_{\lambda=1,2}\int d\Omega
_{k}e^{i\left(  \omega_{m_{H},m_{G}^{\prime}}-\omega_{m_{H}^{\prime},m_{G}%
}\right)  t}\left[  M_{m_{G}^{\prime},m_{H}}^{(\lambda)}\left(  \Omega
_{k}\right)  \right]  ^{\ast}\nonumber\\
&  \times M_{m_{G},m_{H}^{\prime}}^{(\lambda)}\left(  \Omega_{k}\right)
\sum_{\ell,m}(-i)^{\ell}Y_{\ell m}\left(  \Omega_{k}\right)  \left[  Y_{\ell
m}\left(  \mathbf{\hat{R}}_{2}\right)  \right]  ^{\ast}h_{\ell}(k_{0}R_{2}).
\end{align}

We rewrite Eqs. (\ref{ev}) as%
\begin{subequations}
\begin{align}
\dot{b}_{m_{H},m_{G};0}(t) &  =-\gamma b_{m_{H},m_{G};0}(t)-\gamma e^{i\left(
\omega_{m_{H},m_{G}^{\prime}}-\omega_{m_{H}^{\prime},m_{G}}\right)  t}%
\sum_{m_{G}^{\prime},m_{H}^{\prime}}\tilde{G}_{m_{H},m_{G}}^{m_{G}^{\prime
},m_{H}^{\prime}}\left(  \mathbf{R}_{2}\right)  b_{m_{G}^{\prime}%
,m_{H}^{\prime};0}(t),\\
\dot{b}_{m_{G},m_{H};0}(t) &  =-\gamma b_{m_{G},m_{H};0}(t)-\gamma e^{i\left(
\omega_{m_{H},m_{G}^{\prime}}-\omega_{m_{H}^{\prime},m_{G}}\right)  t}%
\sum_{m_{G}^{\prime},m_{H}^{\prime}}\tilde
{G}_{m_{H},m_{G}}^{m_{G}^{\prime},m_{H}^{\prime}}\left(  -\mathbf{R}%
_{2}\right)  b_{m_{H}^{\prime},m_{G}^{\prime};0}(t),
\end{align}
with
\end{subequations}
\begin{align}
\tilde{G}_{m_{H},m_{G}}^{m_{G}^{\prime},m_{H}^{\prime}}\left(  \mathbf{R}%
_{2}\right)  = &  \frac{3}{2}\left(  2H+1\right)  \sum_{\lambda=1,2}\int
d\Omega_{k}\left[  M_{m_{G}^{\prime},m_{H}}^{(\lambda)}\left(  \Omega
_{k}\right)  \right]  ^{\ast}M_{m_{G},m_{H}^{\prime}}^{(\lambda)}\left(
\Omega_{k}\right)  \nonumber\\
&  \times\sum_{\ell,m}(-i)^{\ell}Y_{\ell m}\left(  \Omega_{k}\right)  \left[
Y_{\ell m}\left(  \mathbf{\hat{R}}_{2}\right)  \right]  ^{\ast}h_{\ell}%
(k_{0}R_{2}).
\end{align}
It is possible to integrate over $\Omega_{k}$ to obtain%
\begin{align}
\tilde{G}_{m_{H},m_{G}}^{m_{G}^{\prime},m_{H}^{\prime}}\left(  \mathbf{R}%
_{2}\right)   &  =\frac{\left(  2H+1\right)  }{\left(  2G+1\right)  }%
\begin{bmatrix}
H & 1 & G\\
m_{H}^{\prime} & m_{G}-m_{H}^{\prime} & m_{G}%
\end{bmatrix}%
\begin{bmatrix}
H & 1 & G\\
m_{H} & m_{G}^{\prime}-m_{H} & m_{G}^{\prime}%
\end{bmatrix}
\nonumber\\
&  \times\left[  h_{0}(k_{0}R_{2})\delta_{m_{G}^{\prime}+m_{H}^{\prime}%
,m_{G}+m_{H}}+B\right]  ,\label{14}%
\end{align}
with%
\begin{align}
B &  =\frac{3}{2}\sqrt{\frac{8\pi}{15}}h_{2}(k_{0}R_{2})(-1)^{
m_{G}-m_{H}^{\prime}  }%
\begin{bmatrix}
1 & 1 & 2\\
m_{H}^{\prime}-m_{G} & m_{G}^{\prime}-m_{H} & m_{G}^{\prime}+m_{H}^{\prime
}-m_{G}-m_{H}%
\end{bmatrix}
\nonumber\\
&  \times Y_{2,\left(  m_{G}^{\prime}+m_{H}^{\prime}-m_{G}-m_{H}\right)
}\left(  \mathbf{\hat{R}}_{2}\right)  ,
\end{align}
where $\delta_{a,b}$ is a Kronecker delta. Note that%
\[
\tilde{G}_{m_{H},m_{G}}^{m_{G}^{\prime},m_{H}^{\prime}}\left(  -\mathbf{R}%
_{2}\right)  =\tilde{G}_{m_{H},m_{G}}^{m_{G}^{\prime},m_{H}^{\prime}}\left(
\mathbf{R}_{2}\right)  .
\]
Equation (\ref{14}) can be rewritten as%
\begin{align}
\tilde{G}_{m_{H},m_{G}}^{m_{G}^{\prime},m_{H}^{\prime}}\left(  \mathbf{R}%
_{2}\right)   &  =(-1)^{2(H-G)+m_{G}-m_{H}^{\prime}+m_{G}^{\prime}-m_{H}}%
\begin{bmatrix}
G & 1 & H\\
m_{G} & m_{H}^{\prime}-m_{G} & m_{H}^{\prime}%
\end{bmatrix}%
\begin{bmatrix}
G & 1 & H\\
m_{G}^{\prime} & m_{H}-m_{G}^{\prime} & m_{H}%
\end{bmatrix}
\nonumber\\
&  \times\left[  h_{0}(k_{0}R_{2})\delta_{m_{G}^{\prime}+m_{H}^{\prime}%
,m_{G}+m_{H}}+B\right]
\end{align}
Therefore, for $G=0$ and $H=1$, we can define%
\begin{subequations}
\begin{gather}
\tilde{G}_{m_{H}m_{H}^{\prime}}\left(  \mathbf{R}_{2}\right)  =\tilde
{G}_{m_{H},0}^{0,m_{H}^{\prime}}\left(  \mathbf{R}_{2}\right)  =(-1)^{-m_{H}%
^{\prime}-m_{H}}\left[  h_{0}(k_{0}R_{2})\delta_{m_{H}^{\prime},m_{H}%
}+B\right]  ,\\
B=\frac{3}{2}\sqrt{\frac{8\pi}{15}}h_{2}(k_{0}R_{2})(-1)^{
-m_{H}^{\prime} }%
\begin{bmatrix}
1 & 1 & 2\\
m_{H}^{\prime} & -m_{H} & m_{H}^{\prime}-m_{H}%
\end{bmatrix}
Y_{2,\left(  m_{H}^{\prime}-m_{H}\right)  }\left(  \mathbf{\hat{R}}%
_{2}\right)  ,
\end{gather}
such that%
\end{subequations}
\begin{subequations}
\label{66}%
\begin{align}
\tilde{G}_{11}\left(  \mathbf{R}_{2}\right)    & =\tilde{G}_{-1,-1}\left(
\mathbf{R}_{2}\right)  =h_{0}(k_{0}R_{2})-\sqrt{\frac{\pi}{5}}Y_{2,0}\left(
\mathbf{\hat{R}}_{2}\right)  h_{2}(k_{0}R_{2}),\\
\tilde{G}_{00}\left(  \mathbf{R}_{2}\right)    & =h_{0}(k_{0}R_{2}%
)+2\sqrt{\frac{\pi}{5}}Y_{2,0}\left(  \mathbf{\hat{R}}_{2}\right)  h_{2}%
(k_{0}R_{2}),\\
\tilde{G}_{10}\left(  \mathbf{R}_{2}\right)    & =-\tilde{G}_{0,-1}\left(
\mathbf{R}_{2}\right)  =-\sqrt{\frac{3\pi}{5}}Y_{2,-1}\left(  \mathbf{\hat{R}%
}_{2}\right)  h_{2}(k_{0}R_{2}),\\
\tilde{G}_{01}\left(  \mathbf{R}_{2}\right)    & =-\tilde{G}_{-10}\left(
\mathbf{R}_{2}\right)  =\sqrt{\frac{3\pi}{5}}Y_{2,1}\left(  \mathbf{\hat{R}%
}_{2}\right)  h_{2}(k_{0}R_{2}),\\
\tilde{G}_{1,-1}\left(  \mathbf{R}_{2}\right)    & =-\sqrt{\frac{6\pi}{5}%
}Y_{2,-2}\left(  \mathbf{\hat{R}}_{2}\right)  h_{2}(k_{0}R_{2}),\label{56}\\
\tilde{G}_{-11}\left(  \mathbf{R}_{2}\right)    & =-\sqrt{\frac{6\pi}{5}%
}Y_{2,2}\left(  \mathbf{\hat{R}}_{2}\right)  h_{2}(k_{0}R_{2}),
\end{align}
in agreement with Fu and Berman \cite{fu,corr}.

For atoms on the $z-$axis, $\tilde{G}_{m_{H},m_{G}}^{m_{G}^{\prime}%
,m_{H}^{\prime}}\left(  \mathbf{R}_{2}\right)  $ vanishes unless $m_{G}%
+m_{H}=m_{G}^{\prime}+m_{H}^{\prime}$. For $G=0$ and $H=1$, this implies that
$m_{H}=m_{H}^{\prime}$; excitation exchange cannot transfer excitation between
states differing in $m_{H}$. On the other hand, for $G=1$ and $H=1$,
excitation exchange amplitude $b_{m_{H}=0,m_{G}=0;0}(t)$ can be driven by
amplitudes $b_{m_{G}=0,m_{H}=0;0}(t)$, $b_{m_{G}=1,m_{H}=-1;0}(t)$, and
$b_{m_{G}=-1,m_{H}=1;0}(t)$; that is, there can be a swap of dipole coherence
between excited states differing in their $m$ values.

This research is funded by the Air Force Office of Scientific Research and the National Science Foundation.

\bigskip
\end{subequations}

\end{document}